\documentclass[preprint]{elsarticle}

\usepackage[usenames]{color}
\usepackage{ulem}
\usepackage{amsmath}
\usepackage{mathtools}
\usepackage{changebar}
\usepackage{hyperref}

\newenvironment{revAdded}{}{}

\newcommand{\revDeleted}[1]{}

\newenvironment{rev2Added}{}{}

\newenvironment{reviiiAdded}{}{}

\newcommand{\reviiiDeleted}[1]{}

\newenvironment{revivAdded}{}{}

\newcommand{\revivDeleted}[1]{}

\begin{document}
\title{How to Make a Digital Currency on a Blockchain Stable}
\author[sfc]{Kenji Saito\corref{cor1}}
\ead{ks91@sfc.wide.ad.jp}

\author[wbs]{Mitsuru Iwamura}
\ead{iwamuram@waseda.jp}

\cortext[cor1]{Corresponding author.}

\address[sfc]{Keio Research Institute at SFC, Keio University,
(Murai Lab.) 5322 Endo, Fujisawa, Kanagawa 252-0882 Japan}

\address[wbs]{Graduate School of Business and Finance, Waseda University,
1-6-1 Nishi-Waseda, Shinjuku-ku, Tokyo 169-8050 Japan}

\begin{abstract}
Bitcoin and other similar digital currencies on blockchains are not ideal
means for payment, because their prices tend to go up in the long term
(thus people are incentivized to hoard those currencies), and to fluctuate
widely in the short term (thus people would want to avoid risks of losing
values).

The reason why those blockchain currencies based on proof of work are unstable
may be found in their designs that the supplies of currencies do not respond
to their positive and negative demand shocks, as the authors have formulated
in our past work.

Continuing from our past work, this paper proposes minimal changes to the
design of blockchain currencies so that their market prices are automatically
stabilized, absorbing both positive and negative demand shocks of the
currencies by autonomously controlling their supplies.
Those changes are:
1) limiting re-adjustment of proof-of-work targets,
2) making mining rewards variable according to the observed over-threshold
changes of block intervals, and
3) enforcing negative interests to remove old coins in circulation.
We have made basic design checks
\begin{revAdded}and evaluations\end{revAdded}
of these measures through simple simulations.

In addition to stabilization of prices, the proposed measures may have effects
of making those currencies preferred means for payment by disincentivizing
hoarding, and improving sustainability of the currency systems by making
rewards to miners perpetual.
\end{abstract}


\maketitle

\section{Introduction}

Digital currencies such as BTC on Bitcoin\cite{Nakamoto2008:Bitcoin} blockchain
and Ether on Ethereum\cite{Buterin2014:Ethereum} blockchain are now well-known,
and are accepted means for payment in some countries.
However, these currencies are not ideal as methods of payment because of
instability of their market prices.
Their prices tend to go up in the long term, and to fluctuate widely in the
short term.
Therefore, people are incentivized to hoard those currencies, and would want
to avoid risks of losing values by using them for payment.
Stability of their market prices needs to be achieved if these currencies are
intended to be used as monetary media instead of as investment products.

In the authors' past work\reviiiDeleted{ published as a discussion
paper}\cite{Iwamura2019:BitcoinMonetaryPolicy}\begin{reviiiAdded}
(an improved version of the discussion paper with the same
title)\end{reviiiAdded}, we have
formulated the mechanism of instability by analyzing the economics of those
blockchain currencies based on proof of work.
We have also proposed measures to stabilize the market prices of such
currencies.

The contributions of this paper are refinements of the work as follows:
\begin{enumerate}
\item We have refined the formulation of the economic model
of blockchain currencies based on proof of work, using abstractions closer to
actual currency system designs, and
\item We have refined our proposed measures for stabilization of market prices
of the currencies that would absorb both positive and negative demand shocks.
\end{enumerate}
The proposed measures went through basic design checks
\begin{revAdded}and evaluations\end{revAdded}
by simple simulations.

Our proposals are primarily targeted for Bitcoin and its descendants.
For information of the design of Bitcoin, in addition to the original design
paper\cite{Nakamoto2008:Bitcoin}, we have referred to more up-to-date resources
from the project web page\cite{Bitcoin:ResourcesOnWeb} and the
code\footnote{https://github.com/bitcoin/bitcoin/} itself.

\section{Blockchain Currency based on Proof of Work}

First, we clarify the design of blockchain currencies based on proof of work.

\begin{figure}[h]
\begin{center}
\includegraphics[scale=0.33]{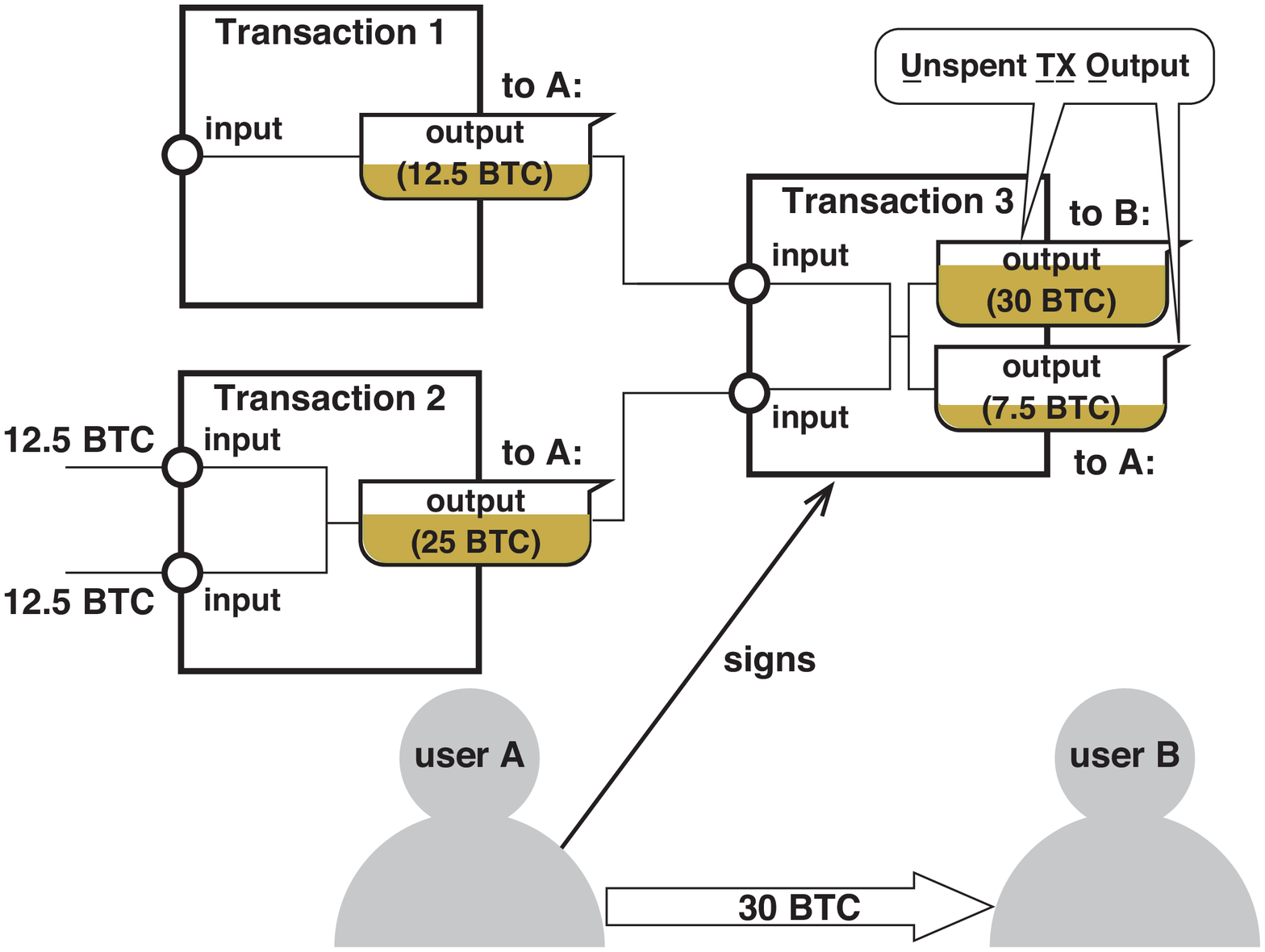}\\
\caption{Transaction Structure of Bitcoin - So-called UTXO Model}
\label{fig-utxo}
\end{center}
\end{figure}

Any participant in the blockchain network may construct and broadcast
(by flooding) a transaction of coins of which they have control.
Figure~\ref{fig-utxo} shows an example of transaction data structure found in
Bitcoin.
A transaction may have a list of inputs each referring to an existing coin as
a container of value (represented as an output of a past transaction),
and a non-empty list of outputs each representing a new coin or null coin (an
output used for applications).
A transaction is digitally signed in each input, which is verifiable with the
public key also included in the input.
The digest of the public key needs to match the digest to which the referred
output is addressed.
This structure is self-contained, and its validity can be verified by anyone.

\begin{figure}[h]
\begin{center}
\includegraphics[scale=0.33]{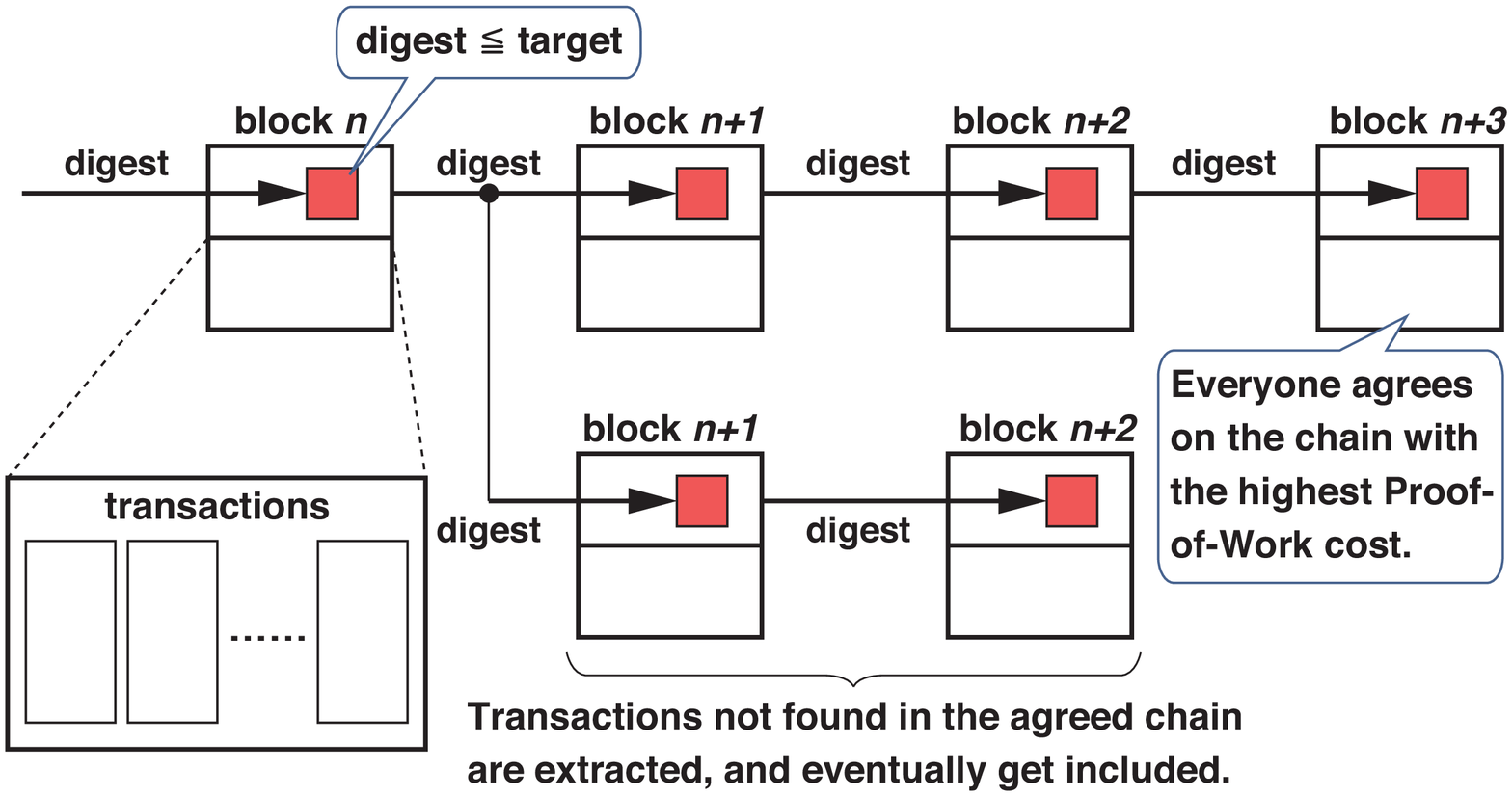}\\
\caption{Overview of Bitcoin Blockchain - Hash Chain with Proof of Work}
\label{fig-blockchain}
\end{center}
\end{figure}

The objective of the blockchain's consensus algorithm
({\em Nakamoto consensus} hereafter) is to maintain a single history
of blocks such that it does not involve any invalid or contradicting
transactions.
An example of the latter is double spending of one digital coin, which cannot
be prevented by the transaction data structure alone.
Presumably, the designer (or designers) of this technology first called it
distributed timestamp server because it is intended to keep track of relative
timing of events.
As Figure~\ref{fig-blockchain} shows, each block contains the cryptographic
digest of the previous block, except the very first block sometimes called the
genesis block.
Such a digest must meet a certain criterion; it needs to be less than or equal
to the pre-adjusted and agreed target stored in or calculated from the block
(we call this structure {\em hash chain with proof of work} hereafter).
Since the digest is calculated by a one-way function whose outputs are evenly
distributed, no one can intentionally configure a block to satisfy the
criterion.
Instead, they need to partake repetitive trials to change the values of some
nonce in the block they are creating until they get a right digest.
Therefore, creation of a block is a probabilistic process.

The necessity of repetitive trials functions as a proof-of-work
mechanism intended to be a protection against falsification.
A transaction itself cannot be falsified unless digital signatures are
compromised.
But it is conceivable to remove some transactions from a past block or to add
fabricated transactions that did not really exist.
If one tries so, the digest of the block is changed and is typically greater
than the target.
Then they would have to retry the proof of work for the block.
This changes the digest stored in the next block, which in turn means that the
digest of the next block is also changed and is typically greater than the
target, and so on.
In short, ones with a malicious intention would have to redo the proof of
work from where they want to change, and outdo the ongoing process of adding
blocks eventually to make the change valid, which has generally been considered
highly difficult.

Such proof of work can also limit the number of proposed blocks at one time.
But there still is a possibility of multiple participants each proposing a
new block at roughly the same time, which may be accepted by different sets of
participants.
Then the hash chain may have multiple ends that are extended independently
from one another, resulting in a fork of the blockchain with multiple (and
possibly, contradicting) histories of blocks.
If this happens, roughly speaking, the longest branch is considered
to be correct.
More precisely, to avoid the case of branches with artificially raised targets,
the branch that is the most difficult to produce is chosen by all participants.
This reflects the total cost cast in the creation of the hash chain branch.
Because of proof of work, any chain branch requires the same cost paid
for its creation when it is tried to be falsified.
In short, Nakamoto consensus tries to enforce that the most difficult chain
branch to falsify is chosen as the single correct
history\footnote{For imperfection of the design of Nakamoto consensus, readers
are referred to a past work\cite{Saito2016:Blockchain} by the first author of
this paper.}.

\section{Market Price Mechanism of a Blockchain Currency
based on Proof of Work}

\subsection{Supply and Demand}

The price of a blockchain currency and proof of work are closely related,
because proof of work is closely related to supply of the coins, and
the price in general is a reflection of supply and demand.

The authors have discussed this in detail in our past
work\cite{Iwamura2019:BitcoinMonetaryPolicy}, and reasoned that the cause
of instability of BTC price is that supply of coins does not respond to demand
shocks.

\begin{figure}[h]
\begin{center}
\includegraphics[scale=0.43]{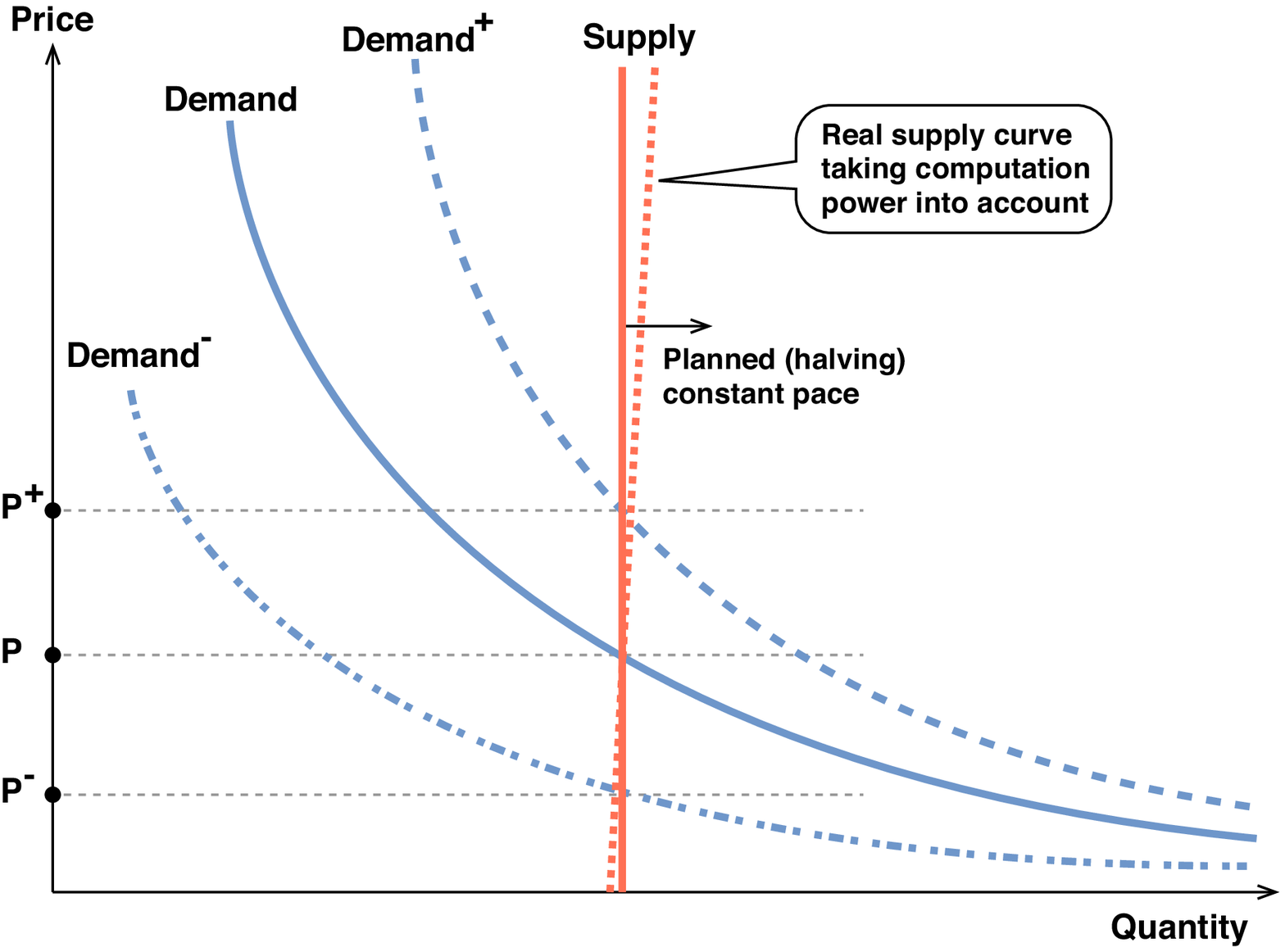}\\
\caption{Supply and Demand of Bitcoin with a Vertical Supply Curve}
\label{fig-supply-demand}
\end{center}
\end{figure}

Figure~\ref{fig-supply-demand} shows the supply and demand curves of Bitcoin.

Usually for any goods (including money),
demand curve slopes downwards from left to right (negatively sloped)
while supply curve slopes positively.
The price is found at the intersection of both curves.
If the demand curve shifts left (demand drops) or right (demand rises),
the supply curve would naturally respond (also shifts left to decreases supply
or right to increase) to ease the shock of demand changes and subsequent
price changes.

On the other hand, BTC coins are only newly supplied as the reward in every
new block.
There just is a rule that the reward is halved every 210,000 blocks.
Therefore the supply curve is vertical
(more precisely, the supply curve is slightly positively sloped,
because when the demand rises and the price goes up, there will be slightly
more supply before the proof-of-work target is adjusted.
When the demand drops and the price goes down, there will be the opposite
effect).
Shifting the demand curve directly affects the price $P$ in the figure.

\subsection{Consequences of Proof of Work}

Here, we present a revised simpler version of the formulation in our past
work\cite{Iwamura2019:BitcoinMonetaryPolicy} to describe the economics of
proof of work.

We start by the number of participants.
Suppose $M$ miners participate.
For simplicity, computation power of all miners are considered equal.
Thus, $M$ is rather the total number of mining units with the same hash rate
(how many digests can be calculated in an interval), and the number can
increase or decrease to represent entry and exit of miners.
\begin{rev2Added}
Further, we assume that there is no monopoly in the mining industry, and
miners' production function does not have externality or
knowledge-intensiveness\reviiiDeleted{, so that the classical, textbook
competition model can substantially be applied to our arguments}.
\end{rev2Added}
\begin{reviiiAdded}

Usually, in the digital world,
a production function is considered virtually free from the law of diminishing
returns because the marginal cost of production is negligible.
However, in the case of mining of a blockchain currency, because the 
proof-of-work target is adjusted to maintain the same production rate of the
currency, adding more production units would at some point yield lower
incremental per-unit returns, showing similar characteristics to productions
in the physical world.
For this reason and with above assumptions, we believe that the classical
competition model (with entry and exit) can be applied to our arguments, as
described below.
\end{reviiiAdded}

Let $\lambda$ be the average number of occurrences of an event that a miner
with one mining unit finds a nonce satisfying the required criterion in an
interval.
We ignore the cost of communication, so that discoveries of a right nonce are
independent events that can globally occur $M\lambda$ times within the
interval.

We consider the average interval between successes $\theta$:

\[\theta = \frac{1}{M\lambda}\]

We assume a cryptographic hash function $H$ with range $R$.
For example, in case of SHA-256, $R = 2^{256}$.
We define the proof-of-work criterion as follows:
the cryptographic digest of a block by $H$ must be less than or equal to
target $G$.
We define target-reachability $g$ as below.

\[g = \frac{G}{R}\]

Conceptually, {\em difficulty} of proof of work is inverse of the
target-reachability $g^{-1}$.

Suppose that one mining unit tries $h$ times in an interval, so that
$h$ represents the average hash rate of a mining unit.
Therefore,

\[\lambda = hg\]

and

\[\theta = \frac{1}{Mhg}\]

Let $V$ be the reward of a successful proof of work.
Then the reward for the winners of the mining competition is fixed as about
$Z$ altogether per interval.

\[Z = \frac{V}{\theta} = VMhg\]

The expected benefit $B_m$ per mining unit per interval is therefore,

\[B_m = \frac{Z}{M} = Vhg\]

If the market value of bitcoin is given as $P$,
the market value of the expected benefit is $PB_m$.

Now, we consider the cost $C_m$ for operating one mining unit.

If $C_m < PB_m$,
more mining units will participate ($M$ is increased).
This in turn will result in shorter $\theta$, and eventually $g$ will be
adjusted with smaller $G$ to maintain the reference $\theta$ (10 minutes in
the case of Bitcoin).
This will result in $B_m$ getting smaller.

Conversely if $C_m > PB_m$,
less mining units will participate ($M$ is decreased).
\begin{rev2Added}
Short-term actualization of this would be that miners turn off their
mining units.
Long-term actualization would be that miners use their mining units to mine
other compatible and profitable coins, or sell their units to others who mine
such coins, or just dispose the units, thereby (partially) leaving from the
mining business of the particular currency.
\end{rev2Added}
Then $\theta$ is prolonged, and eventually $g$ will be adjusted with larger
$G$, resulting in $B_m$ getting larger.

In the long run, the following equilibrium is reached.

\[C_m = PB_m\]

In other form,

\[C_m = PVhg \tag{1} \label{eq-basic}\]

\begin{reviiiAdded}
Some readers may wonder if such an equilibrium is still reached if the price
$P$ goes down to zero.
If $P$ stays at a near-zero level for a long time, the equation requires that
the cost $C_m$ for operating one mining unit is very small.
Please note that we can set $h$ (and subsequently $M$) at an arbitrary level,
so that we can consider very small $h$ (like once in every 10 minutes) and
very large $M$ to start with, where $C_m$ is then very small.
In the case where $P$ is very low, $g$ approaches 1 as $M$ gets smaller and
smaller, and in the end, the equilibrium would be reached.
In reality, if $h$ is that infrequent, a miner can use some idle time of a
computer used for other purposes, so that $C_m$ can actually be negligibly
small ($P$ being near zero is most likely caused by downturn of demands, so
that by the time the near-zero-price situation arises, occurrences of
transactions would become very sparse.
Therefore the cost of verifying transactions can also be ignored).
If $P$ is maintained to be zero for a long time, however, then the equation
cannot hold, as $C_m$ is non-zero albeit it can be negligibly small.
But if that really happens, then there is no economic reason for mining, so
that no mining unit would want to participate, and eventually the blockchain
would stop, in which case there is no point of discussing the equilibrium
anyway.
\end{reviiiAdded}

In the case of Bitcoin, the right side of the equation is contributed as
follows:

\begin{description}
\item[$P$:] Goes up and down mainly by speculation.
\item[$V$:] Starts at 50BTC, and halved at every 210,000 blocks (12.5BTC as of
year 2018).
\item[$h$:] Increases by technological advance.
\item[$g$:] Target $G$ is adjusted every 2,016 blocks to maintain the same
reference value of $\theta$ (10 minutes).
\end{description}

\begin{revAdded}
The left side of the equation $C_m$, the cost for operating one mining unit,
can be roughly modeled as $h$ multiplied by the unit cost of electricity
$C_m^e$, so that $C_m = hC_m^e$.
From the previous equation, we get $hC_m^e = PVhg$.
By dividing both sides with $h$, we get $C_m^e = PVg$.
This suggests that expected gain, $PVg$, will in long term balance with the
unit cost of electricity $C_m^e$.
In a long run, $C_m^e$ decreases while $h$ increases, both by technological
advances, which are possibly motivated by rise of $PVg$.
But their changes are gradual.

In short term,
there is no doubt that $P$ is the driving force of disequilibrium and
subsequent changes in the number of mining units $M$ and then
target-reachability $g$, but $P$ may also be affected by $g$ through people's
expectations, which is affected by $M$.
If $g$ gets halved by $M$ becoming doubled, there is a room for $P$ to get
doubled.
When $V$ gets halved at every 210,000 blocks, if $g$ is unmoved, then $P$ would
need to be doubled to reach the equilibrium (and miners have reasons to want
that to happen).
If it does not happen, miners would need to leave, so that $M$ gets halved, and
in turn, $g$ becomes doubled.

Equation \eqref{eq-basic} gives the long-term target for equilibrium, which we
believe would explain the overall movements of price of bitcoins.
In short term, it seems that the price goes up and down by people's
expectations or operational intentions of the price to go up or down.
This has been made possible because supply does not respond against demand
shocks to begin with --- it is easy to control the price by demands.
We expect that observed wide fluctuations of price in short term would be
improved if a modification is made in the design of the currency system such
that the supply responds against demand shocks.
\end{revAdded}

\section{Measures against Positive Demand Shock}

\subsection{Limited Target Re-adjustments}

To mitigate the effects of positive demand shock, in our past
work\cite{Iwamura2019:BitcoinMonetaryPolicy}, we have suggested
no adjustment to target $G$ unless $\theta$ is very short.

The reason is that we would like to increase the supply of coins by increasing
$Z$ when the demand increases.
To do so, we can only increase $V$ or shorten $\theta$.
What miners can do autonomously through their own decisions is to
increase $M$ (by making more investments), and thereby shorten $\theta$.

But how short is very short, where adjustment of $G$ is eventually needed?
To prevent the blockchain from accidental
forks as described in \cite{Decker2013:BitcoinNetwork}, $\theta$ should have
the minimum limit.
Let it be $\theta_{min}$.
$G$ should be adjusted well before $\theta$ reaches $\theta_{min}$.
So we will adjust $G$ when a certain threshold $\theta'$ is reached
where $\theta' > \theta_{min}$.

\subsection{Target Re-adjustments for Positive Shock}

We propose that $\theta'$ shall be the half of the reference $\theta$ (10
minutes in the case of Bitcoin), where $G$ must get halved.
At the same time, we double $V$ to maintain the same pace of the supply of
coins.

To make this happen, the system needs to refer to real time, which had been
considered difficult for blockchains where timestamps in blocks can be untrue.
But now, we have the concept of 
MTP (Median Time Past; the median time of the 11 blocks in the past) as a
standard measure for the purpose.
We check the average interval between blocks using MTP,
every after 100 blocks, for example.

In the case of Bitcoin, 
because the reference $\theta$ is 10 minutes,
100 blocks takes about 17 hours.
When $\theta$ is halved, it can be detected within a half day or so.

\section{Measures against Negative Demand Shock}

\subsection{Two-fold Propositions}

We propose the measures to mitigate the effects of negative demand shock in
two fold:
\begin{enumerate}
\item No adjustment to target $G$ unless the maximal tolerable average interval
$\theta_{max}$ is reached.
\item Continuous removal of coins in circulation so that less total supply
of coins can be achieved, not only new supply.
\end{enumerate}

For the former, we propose the following adjustment rule for $G$:
when 20\% of $M$ is observed to have left
($\theta$ is observed to be $\frac{5}{4}$ times longer than the reference
value), make $G$ $\frac{5}{4}$ times larger.
At the same time, make $V$ 80\% of the reward at the time to maintain the
same pace of supply of coins.
This adjustment will likely to occur within a couple of days after a negative
demand shock.
We think that this asymmetry in the rules against positive and negative demand
shocks is necessary because time measured with block creations is prolonged
as $\theta$ gets prolonged.

For the latter above, we must invent something new, because blockchain currency
protocols usually include the currency supply rule, but does not have a
currency absorption or write-off rule.

\subsection{Negative Interests}

In our past work\cite{Iwamura2019:BitcoinMonetaryPolicy}, we have suggested
an implicit inflation target.
But in this paper, we propose an implementation of negative interests, and
no halving rule for $V$.

Let us consider {\em coinage eras}, where each era is, for example,
consecutive 2016 blocks.
Hereafter in this section, time is measured by the coinage eras.

Let $Z_t$ be the amount of produced coins at time $t$ (if $t$ is the current
coinage era, $Z_t$ denotes the amount of produced coins in the era so far).

We define a depreciation rule so that effective values of coins are depreciated
as time elapses.
This depreciation can be like 1\% of the original value as of production of
the coin, after every coinage era elapses.
Let the amount of depreciation of coins produced at time $t$ evaluated at
time $T$ to be $D_t(T)$.

The total supply of coins in circulation $S(T)$ at time $T$ is, then
\[S(T) = \sum_{t=0}^T(Z_t - D_t(T))\]
which is expected to remain constant except during positive or negative demand
shocks.
$D_t(T)$ can be a function like below, in the case of 1\% depreciation of the
original at every coinage era.

\[D_t(T) = min(Z_t, Z_t \times 0.01(T - t))\]

\subsection{Implementation for UTXO Structure}\label{sec-impl-utxo}

\begin{figure}[h]
\begin{center}
\includegraphics[scale=0.37]{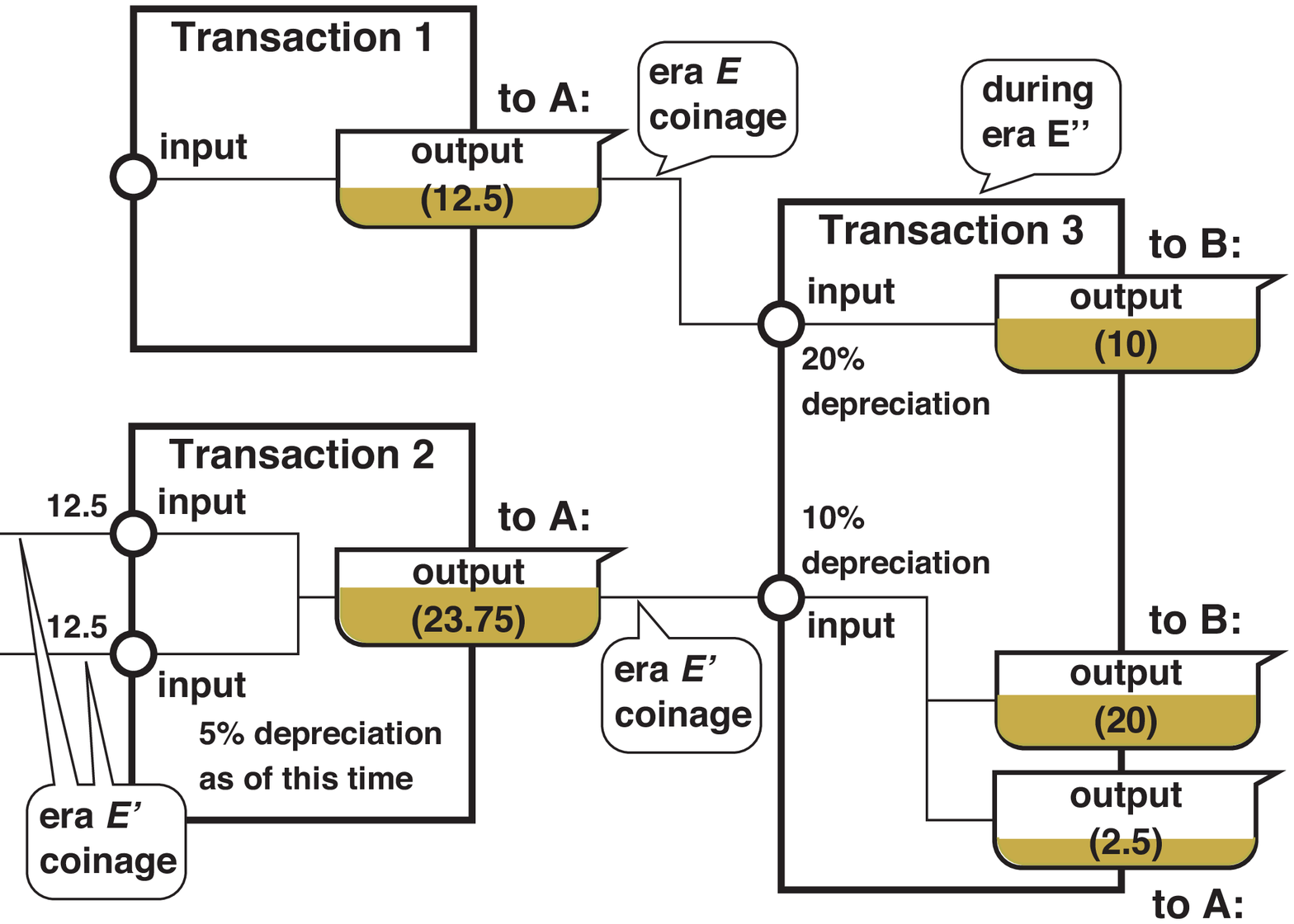}\\
\caption{UTXO Transaction Structure with Depreciation of Coin Values}
\label{fig-ibc-utxo}
\end{center}
\end{figure}

Figure~\ref{fig-ibc-utxo} shows how depreciation is implemented on a UTXO
structure.
In the figure, transaction 1 happened in coinage era $E$,
transaction 2 happened in some coinage era between $E'$ and $E''$,
and transaction 3 is happening in coinage era $E''$.

Because coins produced in different eras have different depreciation levels,
they need to be treated separately in a transaction.
In transaction 3 in the figure, unlike the original design of Bitcoin
transactions, inputs referring to the coin values originated from era $E$ and
era $E'$ are connected separately to different outputs to give total of value
30 to user $B$.
Because coins originated at greater than or equal to 100 eras before do not
exist with the 1\%-per-era depreciation rule
(they are evaluated as value 0, so they cannot be referred to),
at-most 100 kinds of these inputs-outputs connections are expressed in a
single transaction.

A series of inputs-outputs connections originated in an era forms a
{\em coinage era graph} that takes a form of directed acyclic graph.
In the figure, there are two of such graphs, one originated in era $E$ and
another originated in era $E'$.

To evaluate the depreciated coin value of an input, a wallet software to
build a new transaction data or a miner to verify the correctness of the data
does not need to look for all sources and distributions of the
coinage era graph in question.
It just needs to know 1) which era the graph is originated,
2) to which era the transaction being referred to belongs, and
3) the era to which the created transaction is intended to belong.
For example, in the figure, to evaluate the depreciated value at era $E''$ of
the output of transaction 2, one just need to divide 23.75 with 0.95 (because
of 5\% depreciation at transaction 2) to get the original amount, and subtract
10\% from the amount to get 22.5.

With this depreciation rule, when it is near the border of two coinage eras,
a transaction might not make it in the block of the intended era.
If that happens, the transaction data needs to be discarded (or verification
of the transaction would fail anyway because its outputs would be more than
allowed),
and the wallet software needs to recreate the transaction intended for the
new era.
Because of further depreciation in the new era,
the wallet may need to reform the transaction to output the intended amount.

\section{Simulation}

\revDeleted{
To evaluate the effects of the proposed measures, using an artificial market
may look suitable.
However, such an artificial market simulation depends on the human behavior
model being used, and since human reactions against currencies with negative
interests, for example, are not well known yet, we are unsure if we can build
a plausible human behavior model to evaluate our proposals.
It would be easy to produce arbitrary effects.
Furthermore, the law of supply and demand is an established economic theory.
So rather,
}
\begin{revAdded}First,\end{revAdded}
we will make basic design checks of our proposed measures through
simple simulations.
\begin{revAdded}
Then, we will evaluate the measures using an artificial market with simulated
simple human behaviors.
\end{revAdded}

\subsection{Target and Reward Re-adjustments}

\begin{figure}[h]
\begin{center}
\includegraphics[scale=0.45]{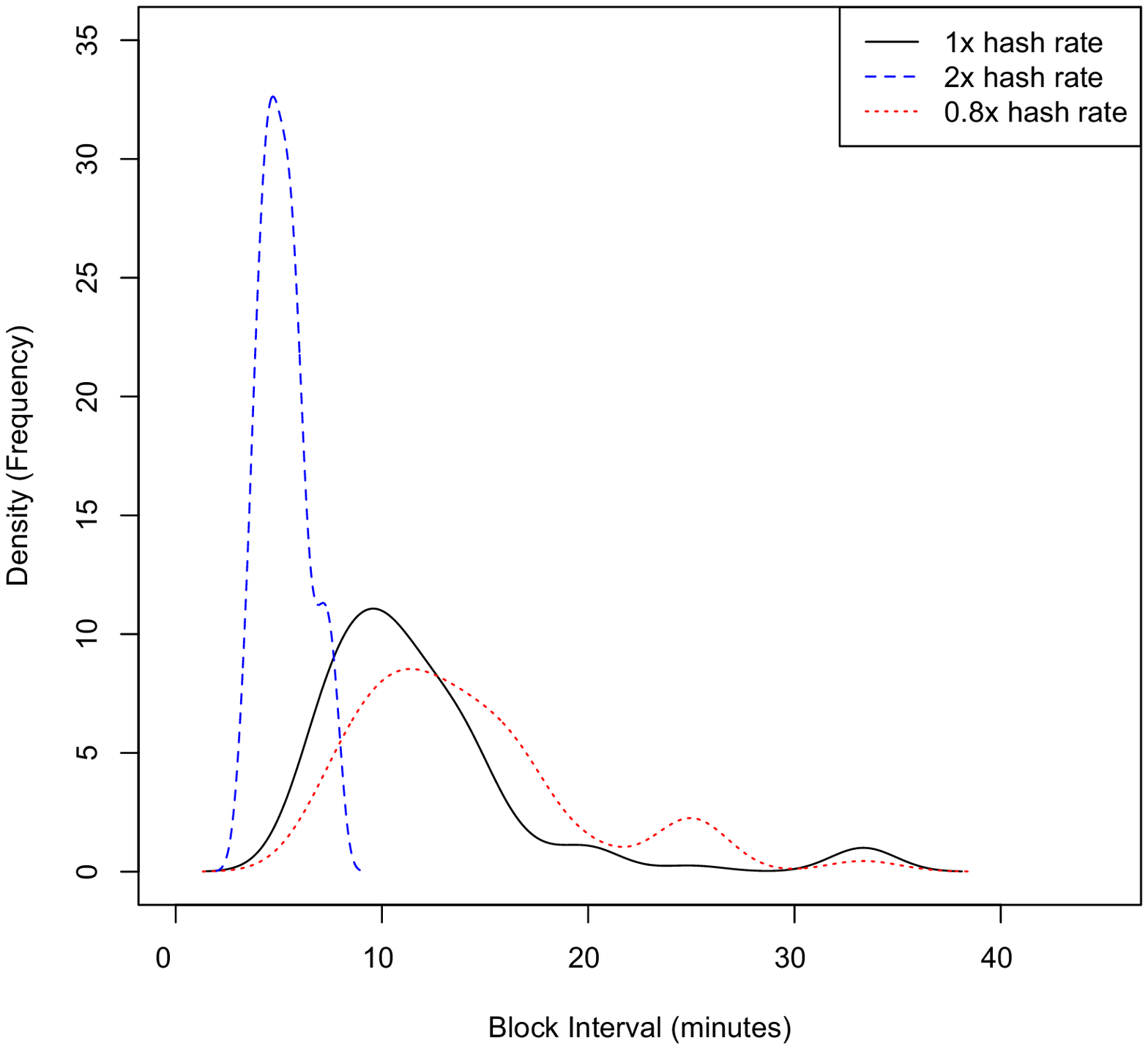}\\
\caption{Distribution of Block Intervals}
\label{fig-block-interval}
\end{center}
\end{figure}

Figure~\ref{fig-block-interval} shows a sample distribution of
block intervals for different total hash rates
($1\times$, $2\times$ and $0.8\times$ of the reference value where $\theta$
equals 10 minutes) to check the
validity of proposed thresholds for target and reward re-adjustments.

We assumed Poisson distribution of events.
We generated 100 random counts of occurrences according to the distribution
in 100 minutes, and plotted them on the graph through
an evaluation of possibility density (frequency of occurrences).
We have tried many iterations, and other iterations show basically the same
shapes.

In the graph, 
$2\times$ of the reference hash rate produces blocks in rather short intervals.
To make this well-longer than $\theta_{min}$,
the currency's network protocol may need to be improved, as suggested in
\cite{Decker2013:BitcoinNetwork}.

On the other hand, $0.8\times$ of the reference hash rate produces blocks in
similar intervals to the reference condition.
This is good in terms of detection of decreased $M$ because it means that the
detection may be possible before users begin to feel that block creations are
getting delayed.
But do not these detections happen too often?
Actual data from Bitcoin blockchain\begin{revAdded}, found at
\cite{BitInfoCharts:BlockTime} for example,\end{revAdded}
shows that this event of the average
interval reaching $\frac{5}{4}$ of the reference (12.5 minutes) is rather
rare (except during \revDeleted{the recent}congestions
\begin{revAdded}observed in late 2017\end{revAdded}).

\subsection{Reduction of Supply}

\begin{figure}[h]
\begin{center}
\includegraphics[scale=0.45]{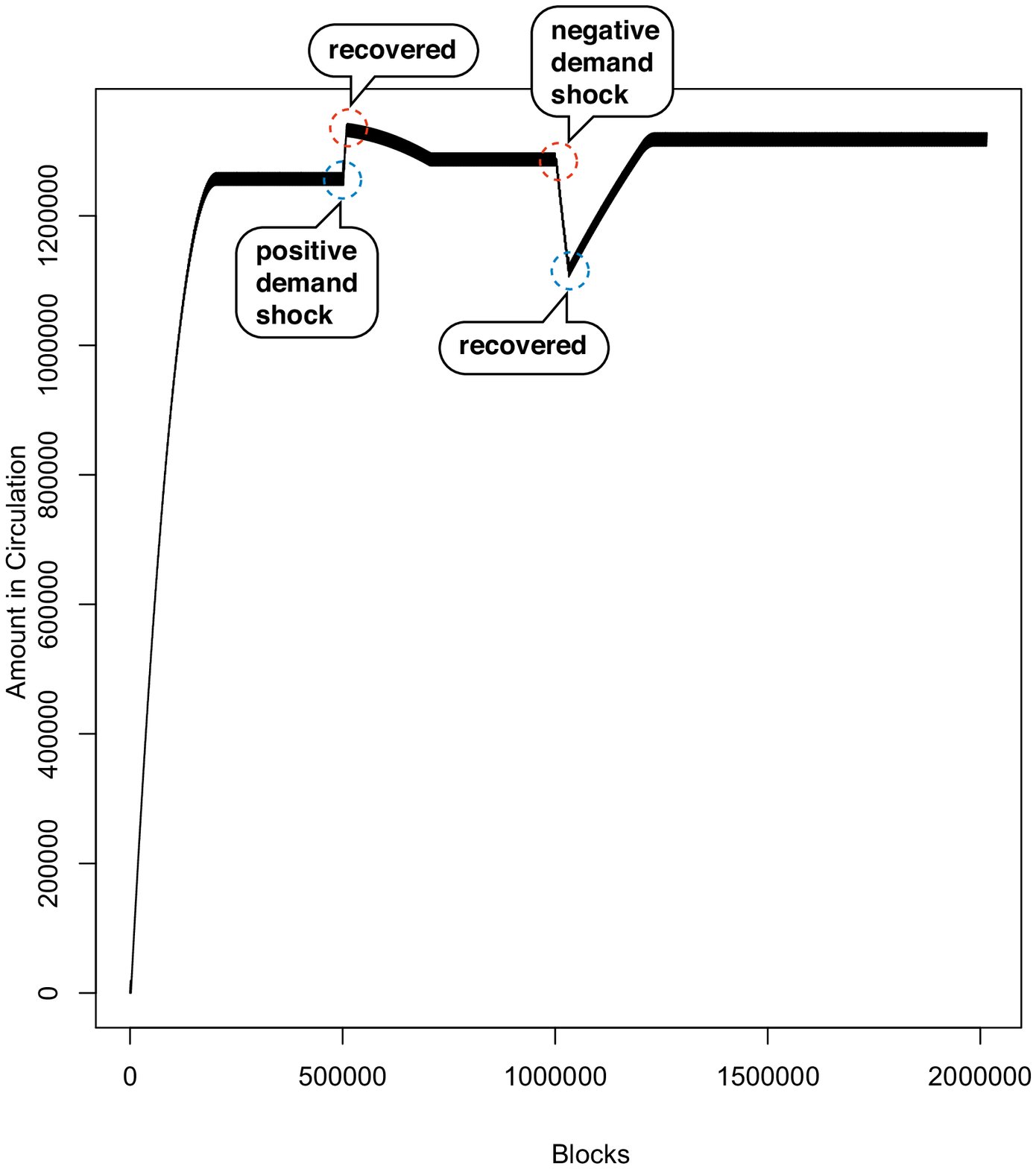}\\
\caption{Simulated Total Amount in Circulation}
\label{fig-sim-negative-interest}
\end{center}
\end{figure}

Figure~\ref{fig-sim-negative-interest} shows a simulation result on the amount
of coins in circulation with depreciation.
In the simulation, we have run the total of 1,000 coinage eras each
containing 2016 blocks (equals to 2,016,000 blocks).
With reference block interval of 10 minutes,
100 blocks (to detect changes in $\theta$) amount to about 17 hours,
2016 blocks (1 coinage era) amount to 2 weeks,
and 100 coinage eras (time for 100\% depreciation) amount to a little less
than 4 years.
But note that the graph is drawn in blocks, not in real time.
In reality, the intervals between blocks may fluctuate largely.
The reward for block creation starts at 12.5, and is adjusted through observed
demand shocks injected artificially during the simulation.

A positive demand shock is injected at block 500,000, which is recovered
around block 506,000, and a negative demand shock is injected at block
1,000,000, which is recovered around block 1,030,000.

Note that slight increase in the supply is caused by offsetting the doubling
rule ($V = V \times 2$) with taking\begin{revAdded}-\end{revAdded}80\% rule
($V = V \times 0.8$) for three
times (amounts to 0.512 of $V$ at the beginning of offsetting).
In reality, supply is also adjusted by autonomous fluctuation of $\theta$.

It shows that the supply of coins remains fairly constant even after positive
and negative demand shocks and their recoveries by the market price mechanism.

\begin{revAdded}

\subsection{Artificial Market}\label{sec-artificial-market}

We call the currency with the proposed measures {\em Stabilized Coin}
hereafter, and compare it with Bitcoin through simulations over an artificial
market we define as follows.

\subsubsection{Model}

\paragraph{Agents} $\:$

We start with defining agents.
We define a set of speculators $U_s$ where $|U_s| = N_s$ and
a set of common traders $U_r$ where $|U_r| = N_r$.
Speculators and common traders are collectively called {\em traders} hereafter.
Each trader $u \in U_s$ or $u \in U_r$ is associated with
market-sensitivity $u^s$ where $0 < u^s \leq 1$, evenly distributed among
the traders in each set.
They are also associated with balance of the coins being simulated (either
Bitcoin or Stabilized Coin, depending on the specific simulation) $u^b$, and
reference price of the currency $u^p$ initialized as price $P$ at the
beginning of each simulation.
In addition, a speculator is associated with previous reference price $u^{pp}$
($u^p = u^{pp}$ to start).
We assume that they have unlimited supply of fiat currency to purchase the
coins, but actual spending of the fiat currency is regulated by probabilities
of making orders and ordering amounts, which are set proportional to $u^s$.

We define a set of mining units $U_m$ where $|U_m| = M$.
$U_m$ is collectively associated with balance of the coins $U_m^b$.

\paragraph{Market} $\:$

We define a daily market $<U_s, U_r, U_m, E>$ where $E$ is an exchange, at
which ordering to buy or sell coins, clearing the orders, and updating the
price $P$ is made once everyday.

For simplicity, miners $U_m$ is purely a supplier of the coins, always trying
to clear $U_m^b$ by selling all coins they have.

Trader $u \in U_s$ or $u \in U_r$ makes an order to buy
$u^b \times u^s \times 0.2$ amount of coins by probability $p$, and to sell
the equal amount of coins by probability $1 - p$
(for random market movements).

In addition, speculators and common traders behave differently as follows:

\begin{itemize}
\item Speculator $u \in U_s$ makes an order to sell $u^b \times u^s \times 0.8$
amount of coins if the price $P$ is greater than
$u^p \times (1 + \frac{u^s}{2})$, and to buy the same amount of coins if $P$ is
less than $u^p \times (1 - \frac{u^s}{2})$
(they try to make profits by ``buy low, sell high'' strategy).
In either case, $u$ updates the reference prices so that $u^{pp} \coloneqq u^p$
and then $u^p \coloneqq P$.
However, in the case that $P$ is less than
$u^{pp} \times (1 - \frac{u^s}{2}) \times \frac{1}{2}$, $u$ makes an order to
buy the same amount of coins as above, but neither $u^p$ nor $u^{pp}$ are
updated (to save the market from crashing and to continue to make more
profits).

\item Common trader $u \in U_r$ makes an order to buy
$u^b \times u^s \times 0.8$ amount of coins if the price $P$ is greater than
$u^p \times (1 + \frac{u^s}{2})$, and to sell the same amount of coins if $P$
is less than $u^p \times (1 - \frac{u^s}{2})$ (they follow market trends,
ending up buying high and selling low).
In either case, $u$ updates the reference price so that $u^p \coloneqq P$.
\end{itemize}

According to the orders, matching is made at the exchange.
Miners' offering is cleared first, and then the priority is set for smaller
trade volumes.

\paragraph{Price Formation} $\:$

We adopt the simplest price formation model according to
\cite{LeBaron2001:Guide},
\[\Delta{}P_t = P_{t + 1} - P_{t} = \alpha(D_t - S_t)\]
where $P_t$, $D_t$ and $S_t$ denote the price, demand and supply at time $t$
(measured by days),
respectively, and $\alpha$ is the sensitivity of price against the difference
between demand and supply.
In the simulations, $\Delta{}P_t$ is truncated to maintain
$-\frac{P_t}{2} \leq \Delta{}P_t \leq \frac{P_t}{2}$.

In reality, it is most likely that $\alpha$ is not constant.
We adopt this simple model so that we can evaluate the effects of $\alpha$
over stability of the price.

\subsubsection{Parameters}

Where applicable, we use parameter values observed in the real Bitcoin
blockchain and trade markets of BTC.
We start our simulation time at the beginning of a reward era, namely when the
reward became 12.5BTC (the start of the current reward era as of 2018).
The reason behind this choice is that miners would in reality behave actively
to boost the currency price upon reward-halving events, which we would like to
avoid implementing in our model for simplicity.
We would like to set the reward-halving event as late in the simulation time as
possible.

From the information at \cite{BitcoinWiki:Supply},
we have identified that the current reward era started on July 9, 2016.
The market price in USD (US Dollar) of bitcoin on that day is found at
``Market Price (USD)'' section of \cite{BlockchainCom:Charts}, and is
652.00USD.

The estimated transaction volume on that day is found at 
``Estimated Transaction Value'' section of \cite{BlockchainCom:Charts}, and is
216,546BTC.
Parameters have been configured to produce more or less similar daily
transaction volumes, assuming that most bitcoin transactions are trades
instead of payments.

Table~\ref{tab-parameters} shows the list of parameters used for the
simulations.

\begin{table*}[h]
\begin{revAdded}
\begin{center}
\caption{Artificial Market Simulation Parameters}\label{tab-parameters}
{\footnotesize
\begin{tabular}{|r|r|l|}\hline
Constant cost $C_m$&		815&Equation \eqref{eq-basic} is assumed to hold at start\\
Initial price $P$&		652.00&	USD price of BTC on July 9, 2016\\
Initial reward $V$&		12.5&	We try 50, 25, 12.5 and 6.25\\
Constant unit hash rate $h$&	1.0&	Abstract rate\\
Initial target-reachability $g$&		0.1&	Abstract probability\\
Initial number of miners $M$&		1000&	Lower-bound is 100\\
\hline
Total coins at the beginning&	15,750,000&	Total BTC generated before July 9, 2016\\
Number of speculators $N_s$&	500&	Initial 10,000,000 coins evenly distributed\\
Number of common traders $N_r$&	1,000&	Initial 5,750,000 coins evenly distributed\\
Constant probability to buy $p$&0.52&	To offset the tendency for larger supply\\
Constant sensitivity of price $\alpha$&	0.0001&	We try 0.00001, 0.0001 and 0.001\\
Simulation time in days&	3,650&	10 years\\
\hline
\end{tabular}
}
\end{center}
\end{revAdded}
\end{table*}

Equation \eqref{eq-basic} is evaluated every day.
$PVhg$ is calculated and compared with the constant cost $C_m$.
$M$ then increases or decreases by 1\% accordingly towards the equilibrium,
with its lower-bound set to be 100.

\subsubsection{Price Changes}

Figure~\ref{fig-sim-btc-stc-basic} shows the results of simulated price
changes of Bitcoin and Stabilized Coin.

\begin{figure}[h]
\begin{revAdded}
\begin{center}
\includegraphics[scale=0.45]{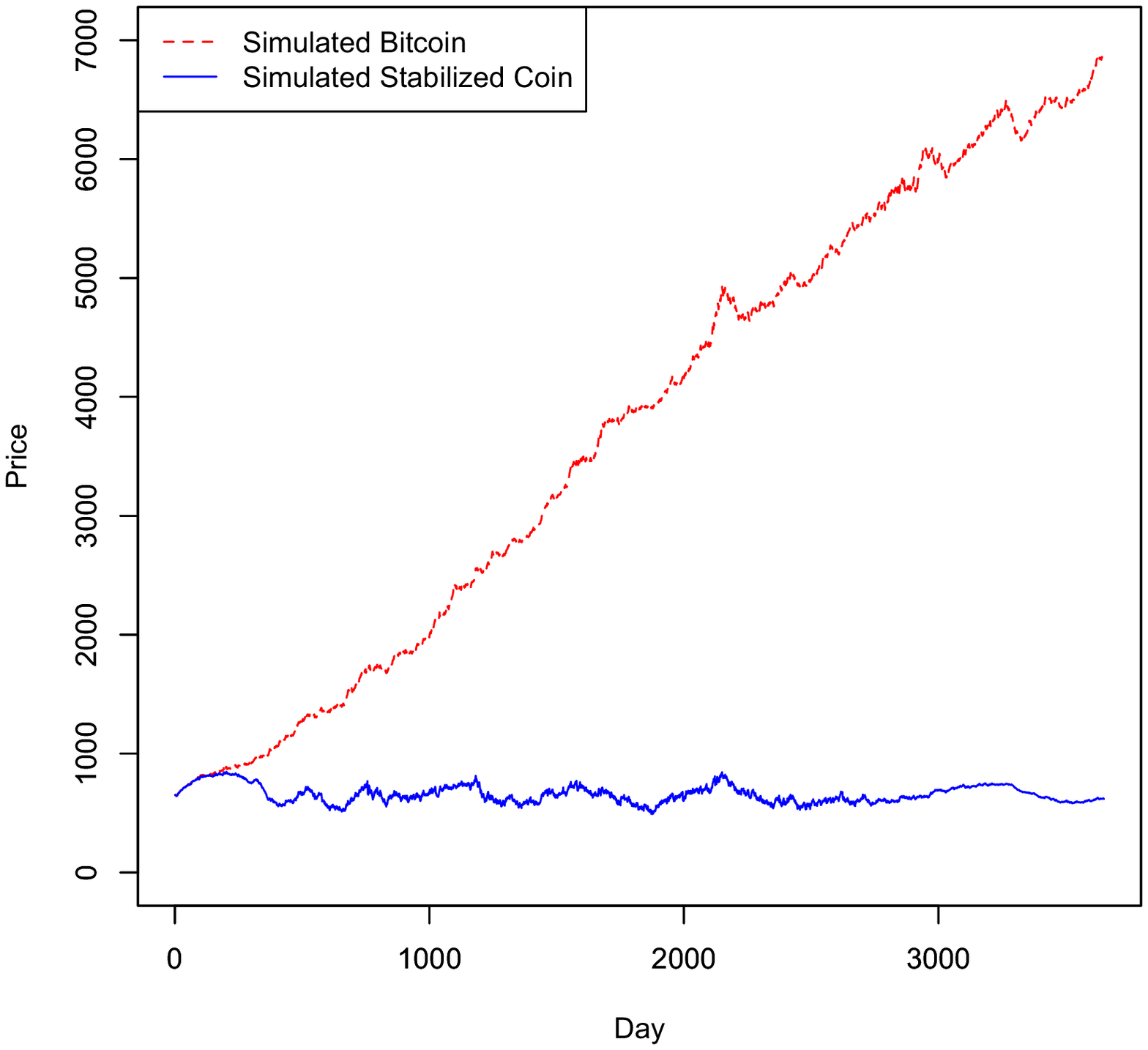}\\
{\footnotesize
\begin{tabular}{|l|lr|lr|}\hline
Simulated Bitcoin&	Min. Price&	642.76&	Max. Price&	6902.94\\
Simulated Stabilized Coin&	&	491.63&			&	846.84\\
\hline
\end{tabular}
}
\caption{Simulated Price Changes of Bitcoin and Stabilized Coin}
\label{fig-sim-btc-stc-basic}
\end{center}
\end{revAdded}
\end{figure}

In reality, Bitcoin price has hit 18,962USD on December 17, 2017 as found at
\cite{BlockchainCom:Charts}, but this simulation shows the maximum price in
10 years as just below 7,000.
We have also seen occasional drops of the price of bitcoin in reality, but
the simulation shows mostly monotonous increase.
By no means, this simulation is a trial to reproduce the events and their
effects observed in reality, but is a demonstration of our model.

The price of Stabilized Coin seems relatively stable in the simulation, but
the fluctuation may seem too big to be used as means for payment
(we will later discuss this in section \ref{sec-sense-price}).

\subsubsection{Transacted Quantity}

Figure~\ref{fig-sim-btc-stc-quantity} shows the results of simulated
transaction quantities of Bitcoin and Stabilized Coin, from the same simulation
as the previous section.

\begin{figure}[h]
\begin{revAdded}
\begin{center}
\includegraphics[scale=0.45]{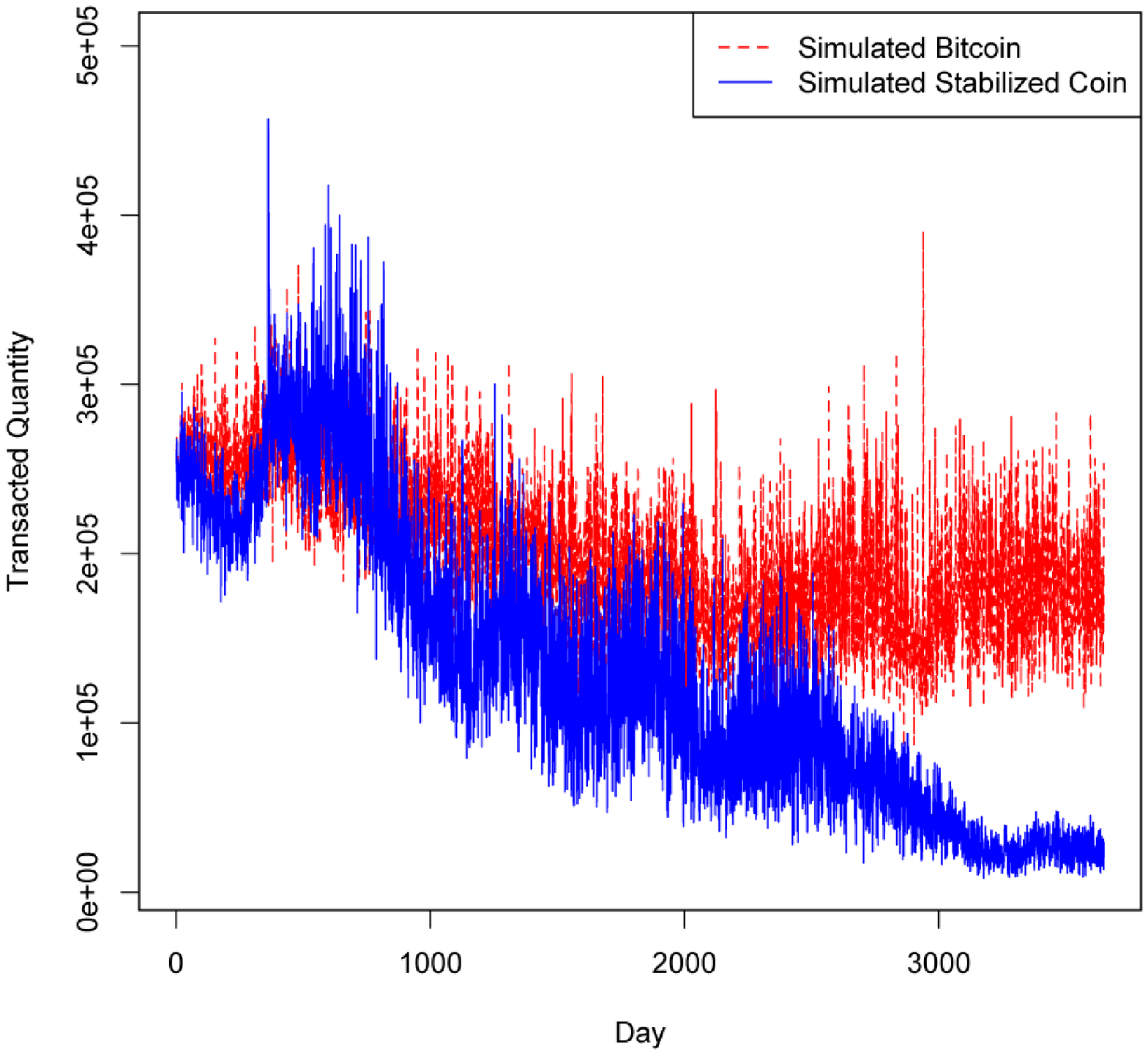}\\
\caption{Simulated Transacted Quantity of Bitcoin and Stabilized Coin}
\label{fig-sim-btc-stc-quantity}
\end{center}
\end{revAdded}
\end{figure}

Those begin at 261,778.7 and 259,159.92, respectively, roughly reproducing the
real quantity of 216,546BTC as found at \cite{BlockchainCom:Charts}.
In the case of the simulated Bitcoin, as long as transacted quantities are
concerned, the simulation seems to be a good reproduction of reality.

\subsubsection{Hash Rate}

Figure~\ref{fig-sim-btc-stc-hashrate} shows the movements of hash rates during
the same simulations of Bitcoin and Stabilized Coin.

\begin{figure}[h]
\begin{revAdded}
\begin{center}
\includegraphics[scale=0.45]{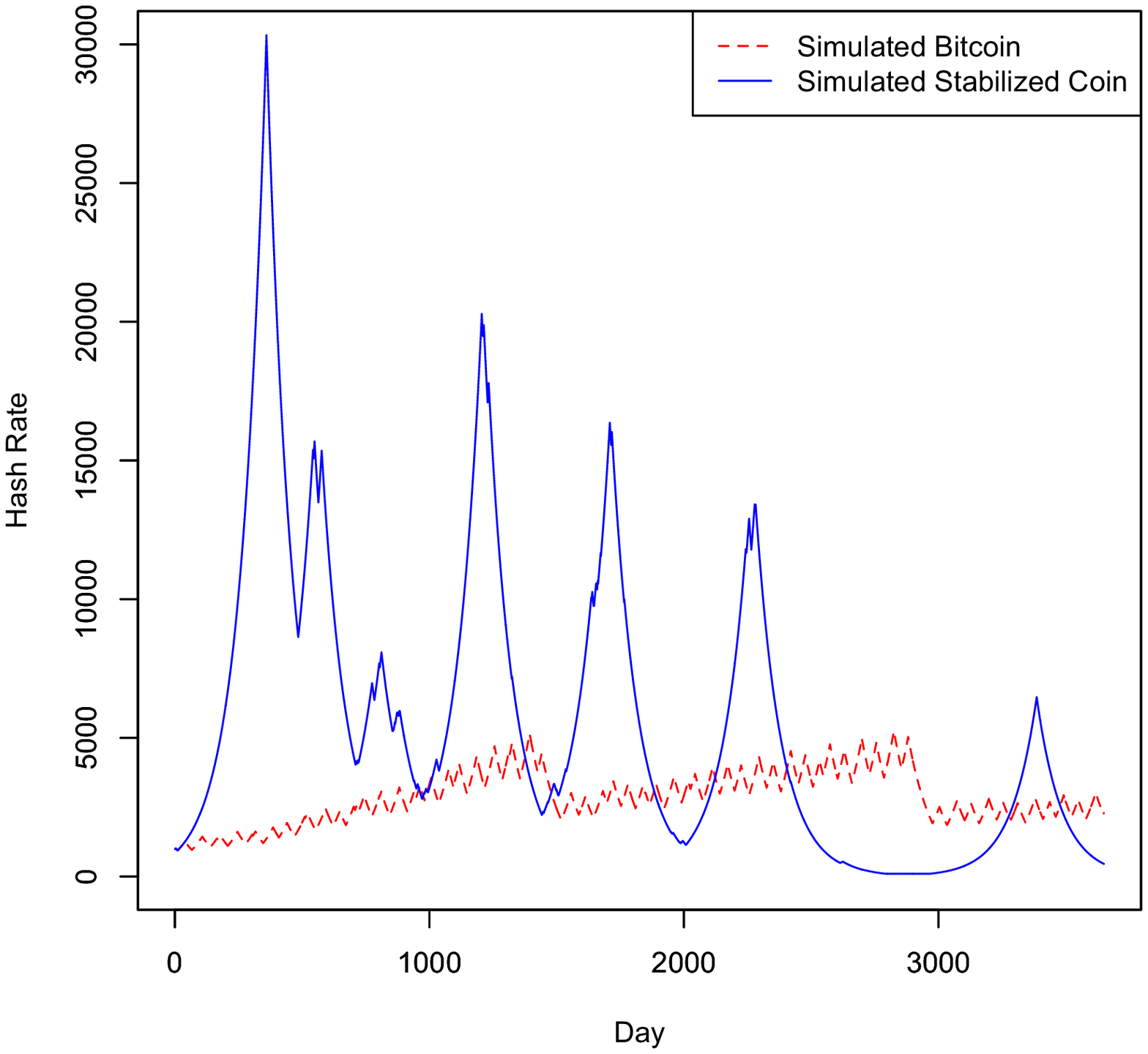}\\
\caption{Simulated Hash Rates of Bitcoin and Stabilized Coin}
\label{fig-sim-btc-stc-hashrate}
\end{center}
\end{revAdded}
\end{figure}

The simulated Bitcoin hash rates do not seem real.
In reality, Bitcoin hash rates grow exponentially.
The reason would be that miners in reality are active traders, and behave like
speculators themselves.
But instead of buying coins while their price is low, they try to produce
coins by mining.

Such behaviors are not included in our model, because it would not fit in the
case of Stabilized Coin, where participating in the production of coins means
taking the role of lowering the currency price.
We did not want to implement different agent behaviors between simulations of
Bitcoin and Stabilized Coin, as it would make fair comparisons difficult.

The simulated Stabilized Coin hash rates goes up and down.
The real hash rate on July 9, 2016 is found at ``Hash Rate'' section of
\cite{BlockchainCom:Charts}, and was 1,580,610TH/s.
Because the initial abstracted total hash rate in this simulation is 1,000,
the highest hash rate in the graph, about 30,000, would amount to
47,418,300TH/s in reality.
The observed highest Bitcoin hash rate so far (as of writing of this
manuscript) was 43,161,058TH/s on June 24, 2018.
So, it seems somewhat feasible, depending on your views.
However, this would raise a more serious question than just feasibility.
The hash rate going up and down relatively rapidly would be a potential
security threat for Stabilized Coin, discussed later in
section \ref{sec-security}.

\subsubsection{Effects of Sensitivity of Price}\label{sec-sense-price}

Figure~\ref{fig-sim-btc-alpha} shows the results of simulated price changes of
Bitcoin where the sensitivity of price $\alpha$ is varied among 0.001,
0.0001 (default in our model) and 0.00001.

\begin{figure}[h]
\begin{revAdded}
\begin{center}
\includegraphics[scale=0.45]{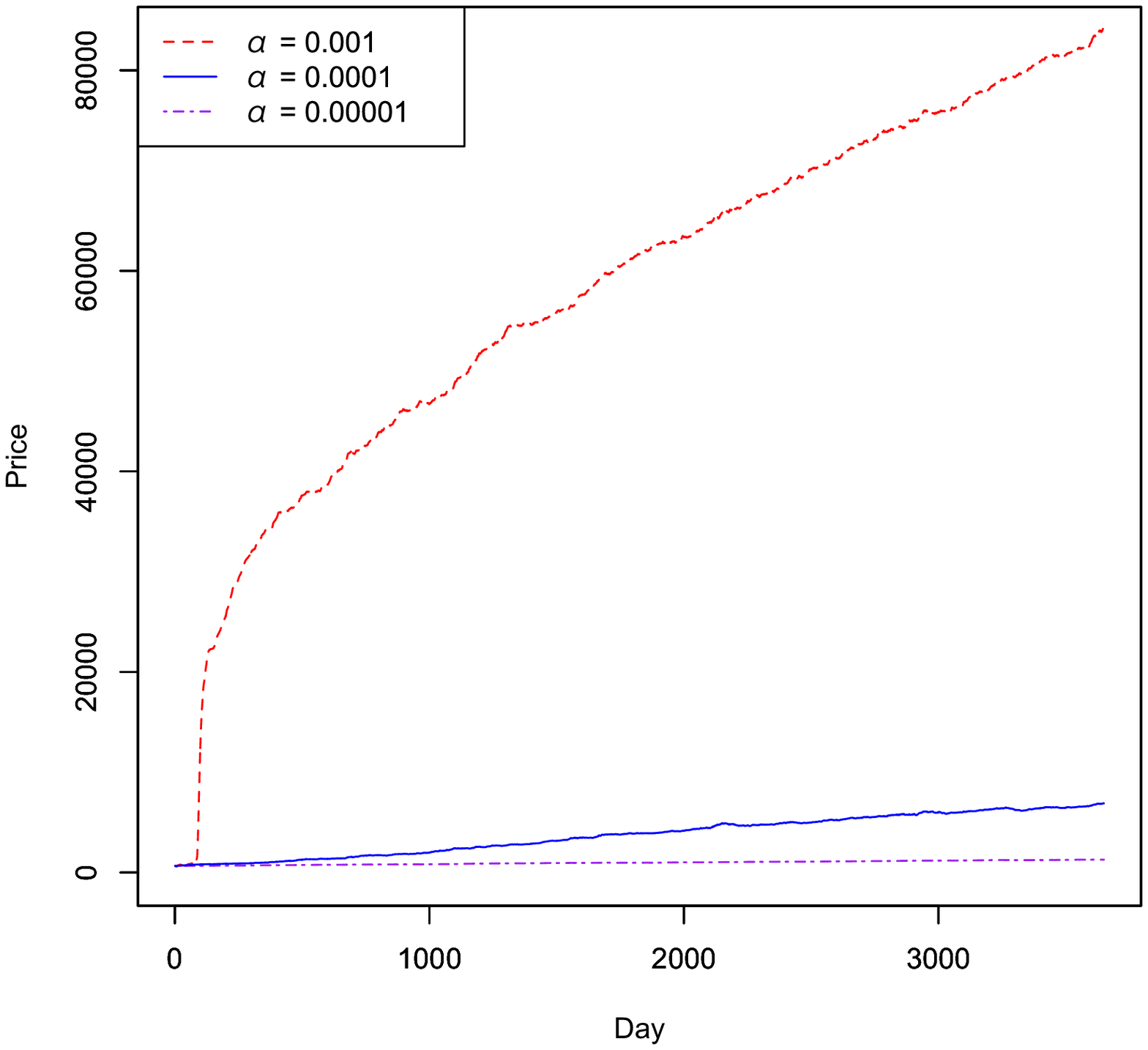}\\
{\footnotesize
\begin{tabular}{|l|lr|lr|}\hline
$\alpha = 0.001$&	Min. Price&	584.30&	Max. Price&	84271.47\\
$\alpha = 0.0001$&	&			642.76&			&	6902.94\\
$\alpha = 0.00001$&	&			651.07&			&	1287.31\\
\hline
\end{tabular}
}
\caption{Effects of Sensitivity of Price $\alpha$ (Bitcoin)}
\label{fig-sim-btc-alpha}
\end{center}
\end{revAdded}
\end{figure}

The effects of $\alpha$ against the simulated Bitcoin prices seem
straightforward.
The larger $\alpha$ is, the higher the price goes.
In the case of $\alpha = 0.00001$, the price seems constant in the graph, but
it actually shows a mostly monotonous increase of the price, which almost
doubles.

Figure~\ref{fig-sim-stc-alpha} shows the results of simulated price changes of
Stabilized Coin where the sensitivity of price $\alpha$ is varied
in the same way.

\begin{figure}[h]
\begin{revAdded}
\begin{center}
\includegraphics[scale=0.45]{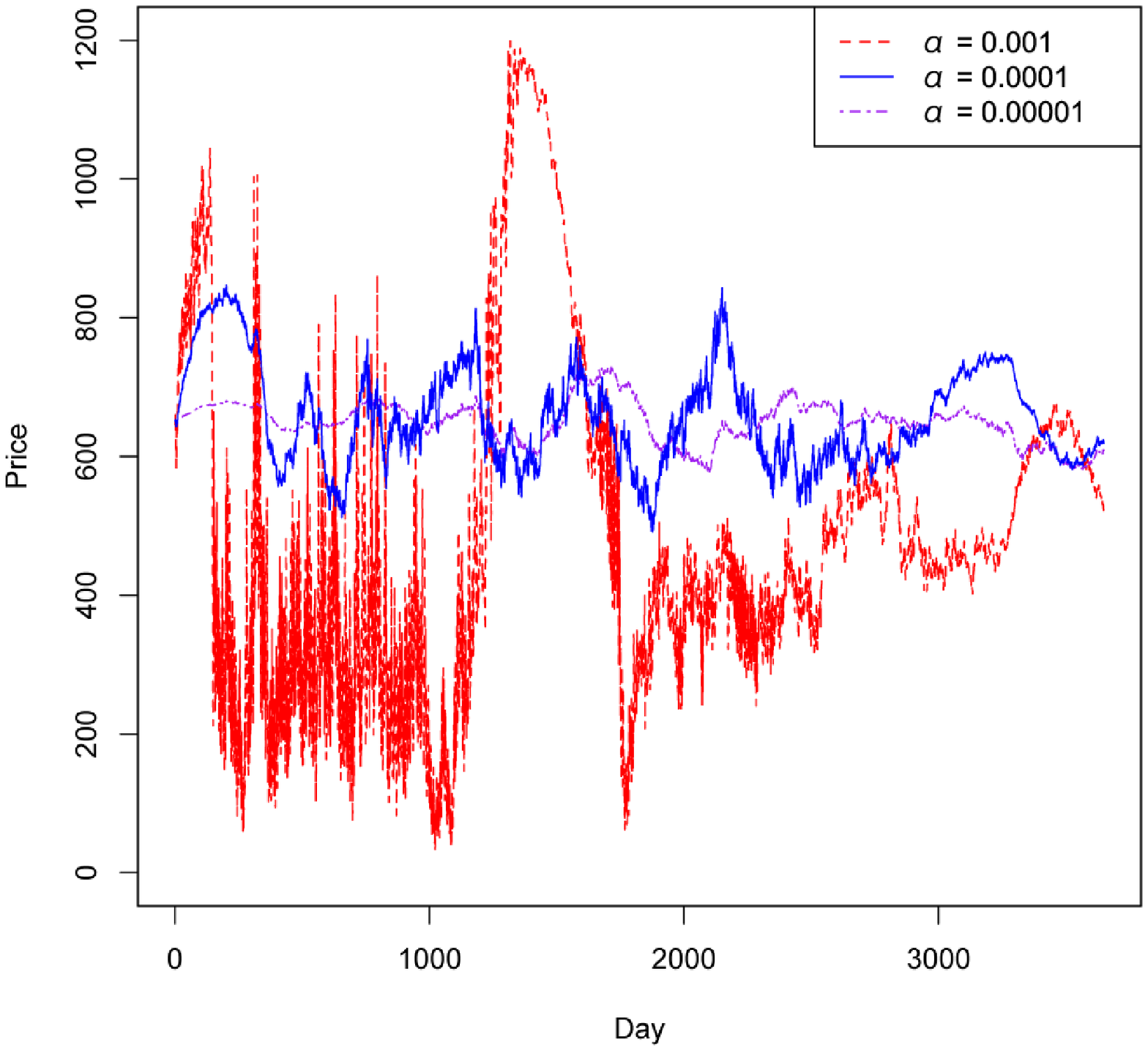}\\
{\footnotesize
\begin{tabular}{|l|lr|lr|}\hline
$\alpha = 0.001$&	Min. Price&	31.96&	Max. Price&	1198.52\\
$\alpha = 0.0001$&	&			491.63&			&	846.84\\
$\alpha = 0.00001$&	&			576.46&			&	728.39\\
\hline
\end{tabular}
}
\caption{Effects of Sensitivity of Price $\alpha$ (Stabilized Coin)}
\label{fig-sim-stc-alpha}
\end{center}
\end{revAdded}
\end{figure}

It is clear from the graph that $\alpha$ takes an important role in stability
of the simulated currency price.
The smaller $\alpha$ is, the more stable the price becomes.
But, even in the case of $\alpha = 0.00001$, the price seems to fluctuate
fairly largely.
We need a reference to evaluate whether these levels of fluctuations are
tolerable or not.

Figure~\ref{fig-sim-stc-usd-jpy} shows a comparison with the historical changes
of USD and Japanese Yen (JPY) exchange rate.
We have obtained the historical data from \cite{Macrotrends:USDJPY}, and
extracted the exchange rates of recent 10 years (July 7, 2008 to July 6, 2018).
To put them in the same scale, we compared the historical changes with the
prices of 0.153 Stabilized Coin (where $\alpha = 0.00001$) and
0.151 Stabilized Coin (where $\alpha = 0.0001$).

\begin{figure}[h]
\begin{revAdded}
\begin{center}
\includegraphics[scale=0.45]{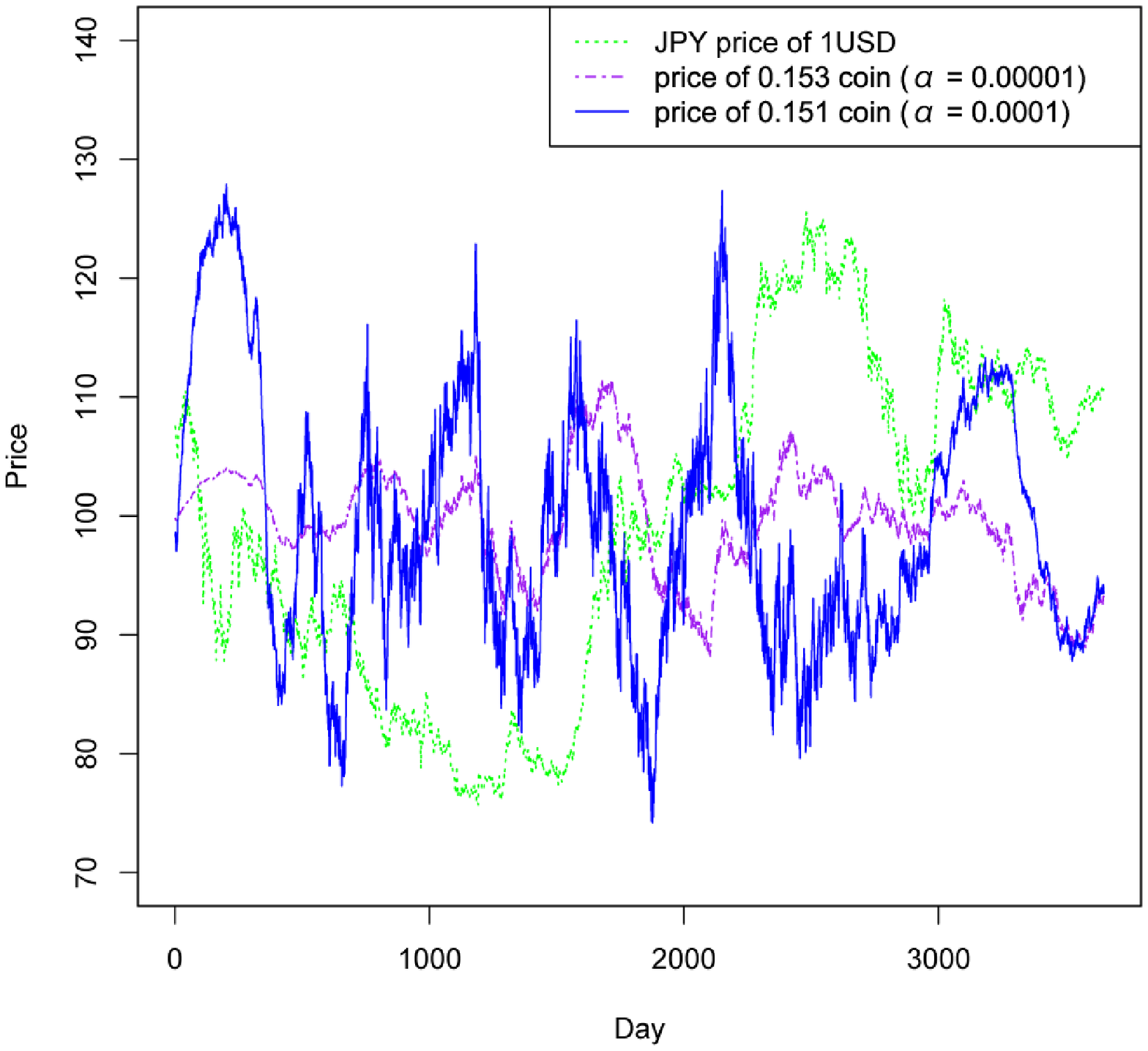}\\
{\footnotesize
\begin{tabular}{|l|lr|lr|}\hline
JPY price of 1USD&	Min. Price&					75.72&	Max. Price&	125.58\\
price of 0.153 coin ($\alpha = 0.00001$)&	&	88.20&			&	111.44\\
price of 0.151 coin ($\alpha = 0.0001$)&	&	74.24&			&	127.87\\
\hline
\end{tabular}
}
\caption{Comparison with Historical Changes of USD-JPY Exchange Rate}
\label{fig-sim-stc-usd-jpy}
\end{center}
\end{revAdded}
\end{figure}

It shows that the price changes of the simulation results are comparable with
the price changes of currencies in real life, although fluctuation is larger
in the case of $\alpha = 0.0001$.

\subsubsection{Effects of Initial Rewards}

Figure~\ref{fig-sim-stc-rewards} shows the results of simulated price changes
of Stabilized Coin where the initial reward for making blocks is varied
among 6.25, 12.5 (default in our model), 25 and 50.

\begin{figure}[h]
\begin{revAdded}
\begin{center}
\includegraphics[scale=0.45]{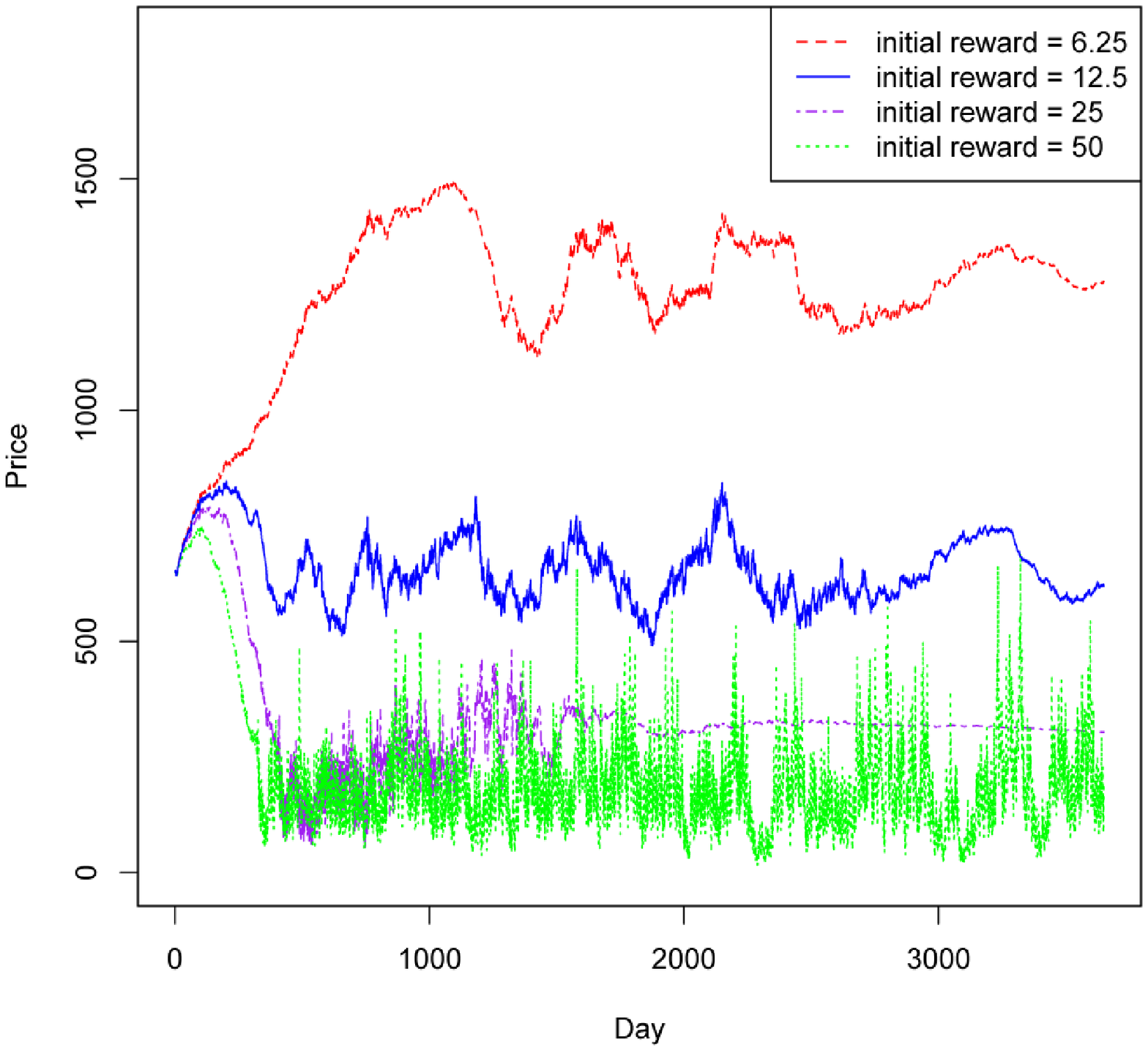}\\
\caption{Effects of Initial Rewards (Stabilized Coin)}
\label{fig-sim-stc-rewards}
\end{center}
\end{revAdded}
\end{figure}

It shows that the smaller the initial reward is, the higher the price tends to
go, because the rewards at block creations represent new supply to the market,
bringing the price of coins down.

In the case of initial reward = 25, the graph shows that the price goes stable
while slowly declining after around day 2,000.
We do not think that this is a success, but instead, it shows limitation of our
simulation model.
In this particular case, the distribution of the coins makes that random
market movements cannot be large enough to cause speculative moves of traders
any more.
\end{revAdded}

\section{Discussion}

\subsection{Responsiveness}

In order for the price stabilization to work, the number of miners $M$ needs
to be quickly responding to demand changes.
But the initial investment cost and sunk cost effects for miners may defer
their entries and exits, although exiting miners have a choice of just turning
(part of) their machines off instead of entirely leaving the business.

Recently, some new blockchain currencies have been started 
by hard forks (backward-incompatible changes of the protocols) from existing
currencies.
It is called a {\em split} of a blockchain.

Ethereum Classic\cite{EthereumClassic:Web}, started in 2016, is probably the
first of such instance, splitted from Ethereum due to a disagreement in
the governance.
Bitcoin Cash\cite{BitcoinCash:Web}, started in 2017, is the first instance of
splitting a new currency from Bitcoin.

Unless further changes are made on the proof of work algorithm, mining these
currencies is compatible with the original blockchain currency.
Therefore, miners may switch back and forth among compatible currencies,
pursuing more profits.
This switch can even be automated.
With this competition among currencies, $M$ can be quickly responding to price
changes according to demand shocks, making the proposed measures more
effective.

\begin{revAdded}
\subsection{Security}\label{sec-security}

However, variance of $M$ would raise security concerns.
A blockchain based on proof of work is protected by the collective cost of
mining.
Bitcoin blockchain is difficult to attack because its hash rate is
exponentially growing, so that existing mining facilities become obsolete
quickly.
If it is not the case, by utilizing available unused facilities, it would be
easier to conduct so-called 51\% attack\footnote{Actually, attack should be
meaningful upon reaching just 50\% of total hash rate, in which case it
becomes uncertain whether the {\em correct} miners would win the race of
creating blocks or not.}
or block withholding attack
(variance of selfish mining\cite{Eyal2018:SelfishMining} to withhold created
blocks while broadcasting contradicting transactions to perform double
spending).
This type of threat is real in many existing cryptocurrencies, 
with actual cases of attacks against currencies called Monacoin and
Bitcoin Gold\cite{CCN:BitcoinGoldAndMonacoin} in May 2018.
This problem seems to be inherent of the design based on proof of work.
\end{revAdded}

\subsection{Applicability to Structures other than UTXO}

We described how to implement depreciation, or negative interests, on a UTXO
structure in section~\ref{sec-impl-utxo}.
But there are blockchain currencies that do not depend on UTXO structures.
Ethereum, for example, maintains the account states in each participating node,
and transactions (encoded state transfers) only are stored in the blockchain.

With this account-state design approach, assuming the rule of 1\% depreciation
over a coinage era we described, 100 different sets of balances are required
for an account to maintain a single currency, for storing the values without
depreciation in the past 100 eras separately.
These values are each negatively weighted as they are depreciated, upon
calculating the effective balances.

\subsection{Deployment}

If we start a new blockchain currency with the proposed measures from its
genesis block, it would have to start with virtually zero price level, and
since the mechanism is intended to avoid speculations, the price level would
remain near zero, not suitable to be used as a monetary medium to represent
values of goods or services.

To avoid such a situation, we may want to utilize a split to employ the
inertia of the original existing currency.
If we start a new currency by splitting it from Bitcoin, for example, the
new currency can start with a useful price level and enough attention,
although how existing coins should be depreciated is a problem to consider.
\begin{revAdded}
In the artificial market simulation in section \ref{sec-artificial-market},
we just implemented depreciation in a straightforward way, not considering
human reactions.
\end{revAdded}

\subsection{Sustainability}

With the proposed measures, the amount of coins in circulation is automatically
maintained to be near constant, and there is no need for external measures to
cap the supply of coins, such as the rule of halving rewards as found in
Bitcoin.
Rewards for block creations to miners can be perpetual, providing a possible
solution to the problem of sustainability in concern.

This means that miners do not need to rely on fees, so we may as well want to
abandon transaction fees altogether.

\subsection{Effects of Depreciation}

In order for the price stabilization to work, the created coins must actually
be supplied into the market.
In our proposal, because coins are depreciated, miners are incentivized to
sell coins they created before they lose values.
The supply is expected to be quick.

But this should be true for all users, not only for miners.
In other words, depreciation should accelerate spending.
The first author of this paper has made a detail study on the effects of
depreciating (and amplifying) digital currencies in
\cite{Saito:2010:BSR:1838759.1838822} before rise of blockchain currencies.
Some of the outcomes of the work may still be applicable.

But a natural question would be, do people prefer currencies that depreciate
over others that do not?
Our answer may be non-intuitive, but {\em yes}.

In our proposal, transactions can be free of charge, as suggested in the
previous section.
Then depreciation can be a beneficial choice for users, because choices are
between always paying transaction fees and possibly avoiding any kinds of it
by spending before depreciation takes place.
(Note that we are discussing digital currencies as media of exchange instead
of investment products.)

\subsection{Effects of Splitting}

Finally, let us consider the case where a currency with proposed measures
itself gets splitted, as it is a possibility for any blockchain currencies
today.

If it happens, it creates a situation where $M$ decreases for both original
and newly splitted currencies.
Thus, both currencies would automatically react as if a negative demand shock
occurred (which may be the case anyway, if the same set of budgets is used for
purchasing both currencies).
In case the new currency took half of the miners, the pace of block creation
is slowed down, but eventually the average interval $\theta$ gets normal, and
the reward $V$ is about halved for both currencies.
If these two currencies compete well, the total coin amount in circulation is
eventually halved for each currency, maintaining the total sum of the amounts
of two currencies to be equal to the amount of the original currency coins in
circulation before the split.
When the process is complete, all coins copied to the new currency upon the
split would have already been depreciated down to zero, and the two currencies
are completely independent from one another in terms of the coins in
circulation.

Therefore, in theory, the split would eventually result in two currencies with
the maintained same level of prices as before and with the same amount in
total in circulation as before.
But how the market in reality reacts against this would need further study.

\section{Related Work}

\subsection{Improvements to Digital Currency Design}

Improvements to the original design of Bitcoin have been proposed many times.
Many of them have taken the forms of alternative coins (altcoins), instead of
research papers.

Among these, one of the most well-known examples is
Litecoin\cite{Litecoin:Web}, from which many other altcoins forked their code.
Litecoin has shown that with basically the same proof of work mechanism as
Bitcoin (although cryptographic hash function in question is different),
average block intervals can be as short as 2 minutes
(Ethereum maintains average block intervals as short as 15 seconds, but with
lots of other improved designs).

Bitcoin Cash was born by a split from Bitcoin, addressing the issue of
improving the throughput by adjustable block sizes.
Many other new currencies have followed the practice of splitting.

Sidechains\cite{Back2014:Sidechains} provide another forms of testing
improvements to blockchains and their currencies.

To our best knowledge, none of these address measures against the negative
demand shock (no means to collect coins in circulation to control supply of
them).

\begin{revAdded}
Another approach towards stabilization of currency prices is
{\em stable coin}\cite{Forbes2018:StableCoin},
which is relatively a new term, but the concept has existed for many years
in the form of electronic money.
The prices of the coins are pegged with fiat currencies.
Tether\cite{Tether:Web} is a well known example of such stable coins.
\end{revAdded}

\subsection{Models of Economics}

Economics of mining has also been modeled in \cite{Cocco2016:BitcoinMarket}
\begin{revAdded}\cite{Cocco2017:CryptocurrencyMarket}\end{revAdded},
where the model is used for simulating the price clearing mechanism in the
market using a virtual order book similar to that presented in
\cite{Raberto2005:PriceFormation}.
\begin{revAdded}
Another trial on understanding economics of Bitcoin through simulations with an
artificial market is found in \cite{Shimada2017:BitcoinPriceDynamics}, using
the terminology and strategies of actual tradings.
\cite{Gandal2014:Competition} makes research on competition among
cryptocurrencies.
\cite{Brandvold2015:BitcoinExchanges} analyzes how Bitcoin prices were
historically formed through exchanges.
\cite{Athey2016:BitcoinPricing} and
\cite{Schilling2018:BitcoinEconomics} each gives a view of Bitcoin Economics.
\cite{Bolt2016:Value} gives such a view on cryptocurrencies in general.
\end{revAdded}

Our model is far simpler
\begin{revAdded}compared with these\end{revAdded}, intended for analysis of
the
problems and devising solutions.
\begin{revivAdded}
The points of our work are twofold: 1) by paying attention to economically
rational behaviors of miners, where mining can be viewed as if it is more or
less like production of goods in firms, we can describe how miners behave
against coin prices in perspective, and 2) by changing the rules for such
behaviors of miners, we can create a negative feedback to the coin prices.

We understand that there may be a different approach, e.g.,
to formulate a total economics such that the coin prices arise endogenously
from the model.
We agree that such a model, if successfully built, has significance.
However, we believe that it is difficult to fulfill our goal with such an
approach, because any price is rational in Bitcoin-like currencies
--- whatever a price may be, once it is predicted that the market would hold
it, then the difficulty target for producing coins will be adjusted
accordingly to the price level, so that any price is effectively an
equilibrium price that balances the supply and demand.
In other words, the prices are determined by the psychological states of those
who are interested in the coins.
It means that the prices can easily be manipulated.
If we could build a model from which the coin prices arise endogenously,
people would still be able to make movements that would break the model.
Instead of working with such a difficulty of handling psychological states of
people, our method is to begin with the view that miners reacting against
prices resemble firms producing goods, and we believe that we have shown that
still it can lead to fulfilling our goal of designing a system that would
make the currency prices stable.

The objective of equation (\ref{eq-basic}) is to formulate behaviors of miners,
instead of
constructing a total economic model for coin prices.
We believe that we have successfully formulated the behaviors of miners, as
our model reflects the reality.

When we see the actual movements of prices and hash rates of Bitcoin\footnote{
\begin{revivAdded}Observable from, for example, https://www.blockchain.com/charts\end{revivAdded}},
we observe general long-term tendencies that the hash rate rises as the
price keeps rising (the all-time tendency of Bitcoin hash rate until September
2018), and the hash rate falls as the price keeps falling (a tendency of
Bitcoin hash rate from September to December 2018 --- a downfall tendency
never observed before).
Miners can freely behave speculatively to try to maintain or raise the values
of the coins they own in the short term, but in the long run, as we have
predicted, their moves seem to conform to the economic rationality of equation
(\ref{eq-basic}).

\end{revivAdded}

However, in order to evaluate the effectiveness of our proposed measures in
more detail,
\begin{revAdded}and to further improve the design,\end{revAdded}
we may want to \revDeleted{use}\begin{revAdded}refine\end{revAdded} the
artificial market methods \begin{revAdded}according to
those\end{revAdded} found in \revivDeleted{these}
\begin{revivAdded}related\end{revivAdded} work
\begin{revivAdded}above\end{revivAdded}
ourselves\begin{revAdded}, with
insights from past work on dynamic behaviors of prices such as
\cite{Beja1980:Prices},
and on agent-based simulations such as
\cite{Raberto2001:Simulation}\end{revAdded}.

\section{Conclusions}

Continuing from our past work, this paper proposed minimal changes to the
design of blockchain currencies so that their market prices are automatically
stabilized, absorbing both positive and negative demand shocks of the
currencies by autonomously controlling their supplies.

\begin{revAdded}
We have evaluated the proposed measures with simple simulations including
those using an agent-based artificial market.
We have shown that resulted price changes are comparable with those of
currencies in real life, suggesting the effectiveness of the measures.
However, more sophistications of the simulations and/or testbed trials would be
needed to further the research.
\end{revAdded}

In addition to stabilization of prices, the proposed measures may have effects
of making those currencies preferred means for payment by disincentivizing
hoarding, and improving sustainability of the currency systems by making
rewards to miners perpetual without need for transaction fees.

\bibliographystyle{elsarticle-num}
\bibliography{stable-currency}

\end{document}